\documentclass[twocolumn,pre,aps]{revtex4-2}
\usepackage{dcolumn}
\usepackage{amsmath}
\usepackage{amsfonts}
\usepackage{amssymb}
\usepackage{calligra}
\usepackage{bm}
\usepackage{graphicx}
\usepackage{float}
\usepackage{subfigure}

\usepackage{stackengine}
\usepackage{natbib}
\usepackage{physics}
\usepackage[]{hyperref}
\usepackage{wasysym}

\begin{document}

\title{Phase Transitions and Critical Behavior in Quasi-One-Dimensional Two-Channel Systems with Quasiperiodic Disorder}

\author{Mohammad Pouranvari}
\email{m.pouranvari@umz.ac.ir}
\affiliation{Department of Solid-State Physics, Faculty of Science,
  University of Mazandaran, Babolsar, Iran.}

\date{\today}

\begin{abstract}
    We investigate the localization properties of a quasi-one-dimensional two-channel system with symmetric and asymmetric onsite energies using the Aubry-Andr\'{e} model. By analyzing the Lyapunov exponent and localization length, we characterize the phase transitions and critical behavior of the system. For the symmetric model, we obtain the phase diagram for the entire spectrum, revealing mobility edges between delocalized and localized states. In contrast, for the asymmetric model, we identify a critical line $ \lambda_1^c + \lambda_2^c \approx 0.5 $ marking the phase transition between delocalized and localized states. We also study the effects of the inter-channel coupling $ \tilde{t} $, and observe that increasing $ \tilde{t} $ reduces the delocalized phase space, shifting the transition from $ \lambda_1 = \lambda_2 = 1 $ at $ \tilde{t} = 0 $ to $ \lambda_1^c + \lambda_2^c \approx 0.5 $ at larger $ \tilde{t} $. Furthermore, the phase transition point is found to be sensitive to both $ \tilde{t} $ and the incommensurate modulation parameter $ b $. While the general phase transition behavior is preserved, subtle differences arise for different values of $ b $, indicating a dependence of the phase boundary on both parameters. Using the cost function approach, we calculate the critical potential strength $ \lambda_c $ and the critical exponent $ \nu $, with $ \nu \approx 0.5 $ for the middle of the spectrum in both symmetric and asymmetric models.
\end{abstract}

\maketitle

\section{Introduction} \label{sec:introduction}

The study of localization phenomena in low-dimensional systems has gained significant attention due to its profound implications for material science and condensed matter physics. Localization, a process where particles become confined due to disorder, plays a crucial role in understanding electronic transport in disordered materials. Two key quantities in characterizing localization are the Lyapunov exponent and the localization length. The Lyapunov exponent provides a measure of the rate of exponential divergence or convergence of nearby trajectories, while the localization length describes the extent to which a wavefunction is spatially localized.

The phenomenon of Anderson localization, first proposed by P.W. Anderson in 1958 \cite{PhysRev.109.1492}, highlighted the impact of disorder on the electronic states in solids, demonstrating that sufficient disorder can localize electrons, preventing them from contributing to electrical conduction \cite{lagendijk2009fifty, kokkinakis2024anderson, kolovsky2024effects, micklitz2024topology, bid2024universality}. This groundbreaking work laid the foundation for numerous studies on disordered systems, including investigations into one-dimensional and quasi-one-dimensional systems. In these systems, localization can occur with any infinitesimal amount of disorder, while in three-dimensional systems, there is a critical strength where the system undergoes a phase transition between delocalized and localized phases \cite{RevModPhys.80.1355, markos2006numerical, doi:10.1080/13642819308215292}. However, some one- and two-dimensional models with correlations among their random numbers exhibit a phase transition between delocalized and localized phases \cite{PhysRevLett.82.4062, mirlin1996transition}. Among these models, the Aubry-Andr\'{e} model has gained significant attention. Introduced in 1980 by S. Aubry and G. Andr\'{e} \cite{aubry1980analyticity}, the Aubry-Andr\'{e} model extended these ideas to quasiperiodic systems, revealing a unique phase transition from extended to localized states without the need for randomness. This model has since been a cornerstone for understanding localization in systems with incommensurate potentials \cite{biddle2009localization, dominguez2019aubry,biddle2011localization, wang2021duality, longhi2019metal}.

The Aubry-Andr\'{e} model, celebrated for its simplicity and rich physical insights, serves as an ideal framework for studying quasiperiodic systems. This model is described by the Hamiltonian:
\begin{equation}
    H = \sum_{n}  t (c_{n+1}^\dagger c_n + c_n^\dagger c_{n+1}) + 2\lambda \cos(2\pi b n) c_n^\dagger c_n,
\end{equation}
where $c_n^\dagger$ and $c_n$ are the creation and annihilation operators at site $n$, $t$ is the hopping amplitude, $\lambda$ is the potential strength, and $b$ is an irrational number. The choice of \( b \) introduces a quasi-periodic potential that is neither completely periodic (where the potential repeats at regular intervals) nor purely disordered (where the potential is random), resulting in a complex, incommensurate structure. This interplay between periodicity and disorder leads to a rich variety of localization phenomena, including the transition between extended and localized states, depending on the system's parameters. This characteristic makes the system particularly intriguing, as it supports the emergence of localized states despite the lack of strict periodicity. Although this model is not random, the incommensurate nature of the potential mimics some effects of disorder, particularly in terms of localization behavior. Additionally, some studies incorporate an extra random phase into the argument of the cosine function, further enhancing the complexity of the system's behavior; however, this work does not include such a modification. The model exhibits a notable phase transition from extended to localized states as the potential strength $\lambda$ is varied. When $\lambda < t$, the states are extended, whereas for $\lambda > t$, the states become localized. This model is a powerful tool for investigating localization phenomena and understanding the effects of quasiperiodicity on electronic properties \cite{liu2020generalized, PhysRevB.105.174206, mckinstry2024localization, bandyopadhyay2024probing, liu2024quantum, liang2024quantum, liu2024emergent, guo2024multiple}.

The localization length $\xi$ is a crucial parameter in understanding the transport properties of disordered systems. It quantifies the extent to which a wavefunction is spatially localized due to disorder. In the context of electronic transport, the localization length is directly related to the conductance $G$ of a system. According to the scaling theory of localization, in one-dimensional and quasi-one-dimensional systems, the conductance decreases exponentially with the length $L$ of the system, $G \sim e^{-L/\xi}$ \cite{PhysRevLett.47.1546, PhysRevLett.42.673}. This behavior highlights that for systems where $L \gg \xi$, the conductance is significantly suppressed, indicating strong localization. Conversely, when $L \ll \xi$, the system behaves as if it is in an extended phase with higher conductance. The characterization of phase transitions between localized and delocalized phases is thus intricately linked to the behavior of the localization length. Critical parameters such as the critical potential strength $\lambda_c$ and the critical exponent $\nu$ can be determined through scaling analyses that involve $\xi$ \cite{doi:10.1080/00018736700101265, BKramer_1993}. These analyses reveal insights into the nature of the phase transition, distinguishing between different universality classes of disordered systems. In the AA model, the critical potential strength is $\lambda_c=1$, and the critical exponent for the localization length is $\nu=1$.

The study of two-channel and multi-channel systems has also gained traction, with works exploring how inter-channel coupling and differing potential strengths influence localization properties and phase transitions. Multi-channel Aubry-Andr\'e models reveal complex phase diagrams that exhibit various phases and transitions, allowing researchers to explore how additional channels affect localization and delocalization conditions. They facilitate the investigation of quasiperiodic potentials, leading to the emergence of single-particle mobility edges, which separate localized and extended states. Additionally, multi-channel systems enable the examination of dimensionality effects, simulating higher-dimensional phenomena in lower-dimensional setups. Experimental realizations with ultracold atoms in optical lattices provide a platform for observing these intricate dynamics, contributing valuable insights into fundamental quantum mechanics and novel quantum states.

Several studies have explored localization phenomena in systems similar to the one we investigate. Rossignolo and Dell'Anna \cite{PhysRevB.99.054211} studied localization transitions in coupled Aubry-Andr\'e chains, identifying an intermediate mixed phase and elucidating the conditions for a uniquely defined mobility edge. Sil et al. \cite{PhysRevLett.101.076803} examined a metal-insulator transition in an aperiodic ladder network with on-site potentials following the quasiperiodic Aubry model, demonstrating exact results for specific electron hopping parameters. Heinrichs \cite{PhysRevB.66.155434, heinrichs2003conductance} studied localization in few-channel disordered wires, focusing on the role of evanescent states and interchain hopping in quasi-one-dimensional systems. Nguyen and Kim \cite{nguyen2012anomalously} numerically investigated the localization properties of a two-channel quasi-one-dimensional Anderson model, finding that increasing the interchain hopping strength enhances localization. Prior et al. \cite{prior2006conductance} explored conductance fluctuations and their impact on localization length in two-dimensional systems.

Our work differs from previous studies in several key ways. We analyze a quasi-one-dimensional two-channel system with distinct onsite energies, introducing complexity through the interplay between intra- and inter-channel dynamics, showing how variations in coupling influence localization and phase transitions, thus modeling more realistic physical systems. We also study the system's behavior at different energy levels, offering a comprehensive view of how the energy spectrum impacts localization and critical parameters. By incorporating these factors, we present a richer phase diagram that captures the influence of both coupling strength and energy dependence.

Our main objective is to analyze the Lyapunov exponent and localization length, identify phase transitions, and determine the critical parameters $\lambda_c$ and $\nu$. Through a systematic study of the inter-channel tunneling strength $ \tilde{t} $ and energy levels, we gain deeper insight into how these parameters influence the localization properties of two-channel systems.

The structure of this paper is as follows: Section \ref{model} describes the model and methodology, including the classification of models and the cost function approach for scaling analysis. Section \ref{results} presents the results and discussion, focusing on the symmetric and asymmetric onsite energy models. Finally, Section \ref{conclusion} concludes the paper with a summary of our findings and suggestions for future work.

\section{Model and Method}
\label{model}
We consider a quasi-one-dimensional model with two channels. Each
channel has intra-channel hopping terms $t$, allowing particles to hop
between adjacent sites within the same channel. There is also an
inter-channel hopping term $\tilde{t}$, allowing particles to hop
between the two channels.

\begin{figure}
\centering
\includegraphics[width=\linewidth]{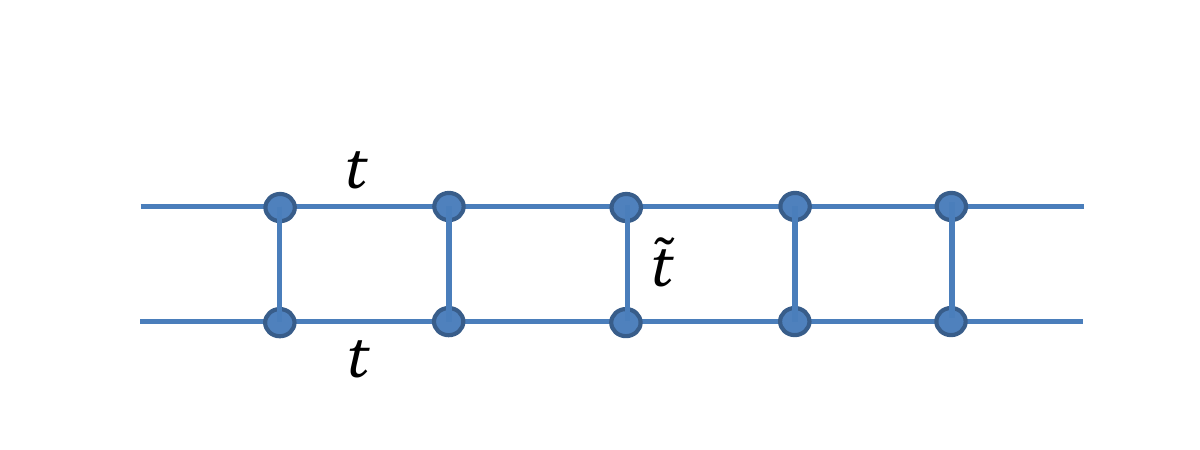}
\caption{Schematic illustration of the two-channel system analyzed in this study. The upper channel and lower channel are coupled through $ \tilde{t} $, while each channel exhibits quasiperiodic onsite potentials.}
\label{fig:model}
\end{figure}

The Hamiltonian of the system is given by:
\begin{eqnarray}
  H &=& \sum_n \epsilon_n^1 a^{\dagger}_n a_n + \epsilon_n^2 b^{\dagger}_n b_n \nonumber \\
  &+& t \sum_n (a^{\dagger}_n a_{n+1} + b^{\dagger}_n b_{n+1}) + \text{h.c.} \nonumber \\
  &+& \tilde{t}\sum_n  a^{\dagger}_n b_n + \text{h.c.},
\end{eqnarray}
Here, $a^{\dagger}_n$ and $a_n$ are the creation and annihilation
operators for a particle at site $n$ in the first channel,
respectively. Similarly, $b^{\dagger}_n$ and $b_n$ are the creation
and annihilation operators for a particle at site $n$ in the second
channel. The term $\epsilon_n^1$ represents the
onsite energy for particles in the first channel, and $\epsilon_n^2$ represents the onsite energy for particles in the
second channel (See Fig. \ref{fig:model})

The onsite energies $\epsilon_n^1$ and $\epsilon_n^2$ follow the
Aubry-Andr\'{e} model:
\begin{equation}
  \epsilon_n^1 = 2 \lambda_1 \cos(2 \pi b n )
\end{equation}
\begin{equation}
  \epsilon_n^2 = 2 \lambda_2 \cos(2 \pi b n )
\end{equation}
where $\lambda_1$ and $\lambda_2$ are the strength parameters of the
onsite potential for the first and second channels, respectively. $b = (\sqrt{5} - 1) / 2$ is an irrational
number, chosen to introduce quasiperiodicity into the system.

To calculate the Lyapunov exponent $ \gamma $, we start with the matrix equation:
\begin{equation}
  \begin{pmatrix}
    \psi_{n-1}^1 + \psi_{n+1}^1 \\
    \psi_{n-1}^2 + \psi_{n+1}^2
  \end{pmatrix}
  =
  T
  \begin{pmatrix}
    \psi_n^1 \\
    \psi_n^2
  \end{pmatrix}
\end{equation}
where $ T $ is the matrix given by:
\begin{equation}
  T = \begin{pmatrix}
    \frac{E - \epsilon_n^1}{t} & -\frac{\tilde{t}}{t} \\
    -\frac{\tilde{t}}{t} & \frac{E - \epsilon_n^2}{t}
  \end{pmatrix}
\end{equation}

To compute the Lyapunov exponent, we first diagonalize the matrix $T$ by setting $\epsilon_n^1 = \epsilon_n^2 = 0$, which corresponds to decoupling the inter-chain coupling terms. This procedure leads to independent channels for wave propagation. We then obtain the eigenvectors $U$ of the diagonalized matrix, and subsequently reintroduce the onsite energies, yielding the modified equation\cite{PhysRevB.66.155434, heinrichs2003conductance}:

\begin{equation}
  U^{-1}
  \begin{pmatrix}
    \psi_{n}^1 \\
    \psi_{n}^2
  \end{pmatrix}
  =
  \begin{pmatrix}
    \tilde{\psi}_{n}^1 \\
    \tilde{\psi}_{n}^2
  \end{pmatrix}
\end{equation}

This leads to:
\begin{equation}
  \begin{pmatrix}
     \tilde{\psi}_{n+1}^1 \\
     \tilde{\psi}_{n}^1 \\
     \tilde{\psi}_{n+1}^2 \\
     \tilde{\psi}_{n}^2
  \end{pmatrix}
  =
  X_n
  \begin{pmatrix}
     \tilde{\psi}_{n}^1 \\
     \tilde{\psi}_{n-1}^1 \\
     \tilde{\psi}_{n}^2 \\
     \tilde{\psi}_{n-1}^2
  \end{pmatrix}
\end{equation}

where $X_n$ is given by:
\begin{equation}
  X_n = \begin{pmatrix}
    \alpha_1 & -1 & \alpha_2 & 0 \\
    1 & 0 & 0 & 0 \\
    \alpha_2 & 0 & \alpha_3 & -1 \\
    0 & 0 & 1 & 0
  \end{pmatrix}
\end{equation}

with
\begin{align}
  \alpha_1 &= \frac{1}{t}(E - \frac{\epsilon_n^1+\epsilon_n^2}{2} -\tilde{t}) \\
  \alpha_2 &= \frac{1}{2t}(\epsilon_n^1 - \epsilon_n^2) \\
  \alpha_3 &= \frac{1}{t}(E - \frac{\epsilon_n^1+\epsilon_n^2}{2} +\tilde{t})
\end{align}

The matrix $M$ is the product of $X_n$ from $n = 1$ to $N$ (the system size):
\begin{equation}
    M = \prod_{n=1}^{N} X_n.
\end{equation}

Using the Landauer formula, we calculate the Lyapunov exponent $\gamma$ as:
\begin{equation}
\gamma = \lim_{N \to \infty} \frac{1}{N} \ln \left( ||M|| \right),
\end{equation}
where $ ||M||^2 = \text{tr}(M M^{\dagger}) $. Since the model does not involve random disorder, no averaging is performed over disorder realizations or eigenenergies.

The Lyapunov exponent ($\gamma$) and localization length ($\xi$) are
intrinsically related in disordered systems. Specifically, the
localization length can be expressed as $\xi =
\frac{1}{\gamma}$. A positive Lyapunov exponent ($\gamma > 0$)
indicates localized states, where wavefunctions decay exponentially,
leading to a finite localization length; in this case, an increase in
$\gamma$ corresponds to a decrease in $\xi$. Moreover, a zero Lyapunov exponent ($\gamma \leq 0$) suggests the presence of
extended states, which are spread throughout the system, resulting in
an infinite localization length ($\xi \to \infty$).

In this study, we calculate $\gamma$ for two distinct models, which
are described in the following section.

\subsection{Classification of Models}
In this study, we classify the quasi-one-dimensional two-channel
system into two distinct models based on the onsite energy parameters
of the channels. This classification allows us to explore the
fundamental differences in the localization behavior and phase
transitions of the system under symmetric and asymmetric potential
landscapes. By examining both identical and differing onsite energies,
we aim to gain comprehensive insights into the role of disorder and
symmetry in the localization phenomena of low-dimensional systems.

Identical Onsite Energies for Both Channels. In the first scenario, we
consider the case where the onsite energies for both channels are
identical, i.e., $\lambda_1 = \lambda_2$. The identical onsite
energies ensure that the potential landscape in both channels is the
same, providing a symmetric environment for the particles.

Different Values for $\lambda_1$ and $\lambda_2$ for the Two
Channels. In the second scenario, we consider different values for the
onsite energy parameters, $\lambda_1$ and $\lambda_2$, in the two
channels. This asymmetry introduces a more complex potential
landscape, reflecting real-world systems where different channels or
components may have varying potential strengths. The difference in
$\lambda_1$ and $\lambda_2$ breaks the symmetry of the system, leading
to richer and more diverse localization phenomena. Analyzing this case
helps us understand how varying potential strengths affect the
localization properties and phase behavior of the system.

\subsection{Using the Cost Function for Scaling Analysis}
\label{costfunction}

The critical exponent $\nu$ associated with the localization length $\xi$ can be determined through scaling analysis using a cost function approach \cite{PhysRevB.102.064207}. Near the critical potential $\lambda_c$, the localization length follows the scaling relation:
\begin{equation}
\xi \sim (\lambda - \lambda_c)^{-\nu}.
\end{equation}

To find $\nu$ and $\lambda_c$, we employ a cost function that measures the quality of the data collapse by minimizing discrepancies between datasets for different system sizes $N$ and potential strengths $\lambda$. The rescaled localization lengths, $\xi / N$, are used to normalize the data. The cost function is defined as:
\begin{equation}
\mathcal{C}_Q = \frac{\sum_{i=1}^{N_p-1} |Q_{i+1} - Q_i|}{\max\{Q_i\} - \min\{Q_i\}} - 1,
\end{equation}
where $Q_i$ are the rescaled localization lengths, and $N_p$ is the total number of points, sorted according to the scaling variable $N (\lambda - \lambda_c)^{\nu}$.

We determine $\lambda_c$ and $\nu$ by minimizing $\mathcal{C}_Q$ through the following steps:
1. Initial Optimization: Broad bounds for $\lambda_c$ and $\nu$ are set, and optimization is performed using a differential evolution algorithm.
2. Refining Bounds: Based on the initial results, the bounds are refined to focus the search on a more relevant region.
3. Multiple Runs: Several optimization runs with different random seeds help explore the parameter space more thoroughly.

The optimal $\lambda_c$ and $\nu$ are selected from the run with the lowest cost function, and successful data collapse is verified by plotting rescaled localization lengths against the scaling variable. Small variations in $\lambda_c$ and $\nu$ may occur between evaluations of the same dataset.

\section{Results and Discussion}
\label{results}
In this section, we present and analyze the results obtained from our investigation of the quasi-one-dimensional two-channel system with symmetric and asymmetric onsite energies. We examine the behavior of the Lyapunov exponent and localization length to understand the localization phenomena and phase transitions in these systems. Our analysis is divided into two parts: the symmetric onsite energy model and the asymmetric onsite energy model.

\subsection{Symmetric Onsite Energy Model}

In the symmetric onsite energy model, where $\lambda_1 = \lambda_2$,
we investigate the behavior of the Lyapunov exponent and localization
length under identical onsite energies for both channels.

\begin{figure}
\centering
\includegraphics[width=\linewidth]{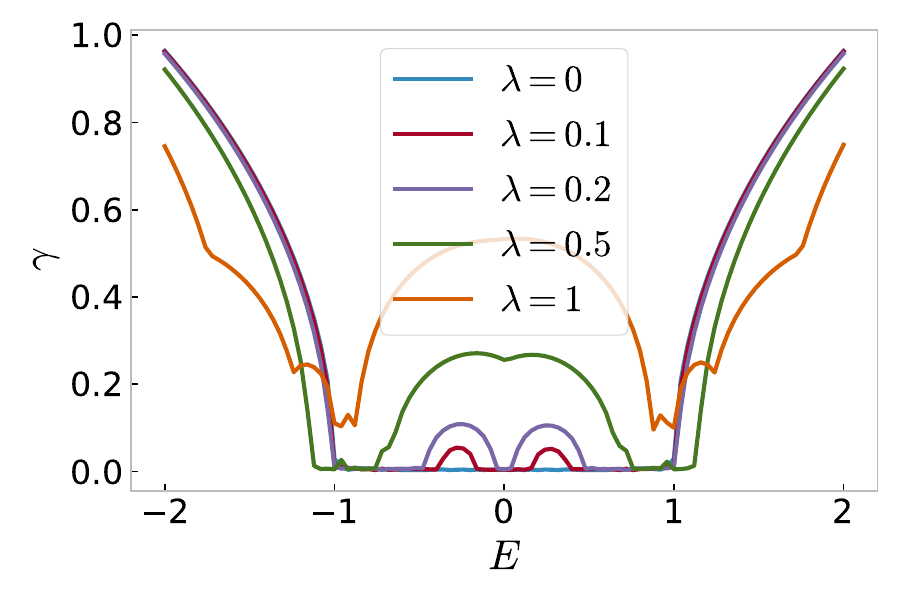}
\caption{Lyapunov exponent $\gamma$ as a function of energy $E$ for
  various values of $\lambda$ in the symmetric onsite energy model. We
  can see distinct behavior of the Lyaponuv exponent at the middle of
  the spectrum and at the tail of it. We set $N=200$, and $t=\tilde{t} = -1$.}
\label{fig:lyapunov_vs_energy_symmetric}
\end{figure}

\begin{figure}
\centering
\includegraphics[width=\linewidth]{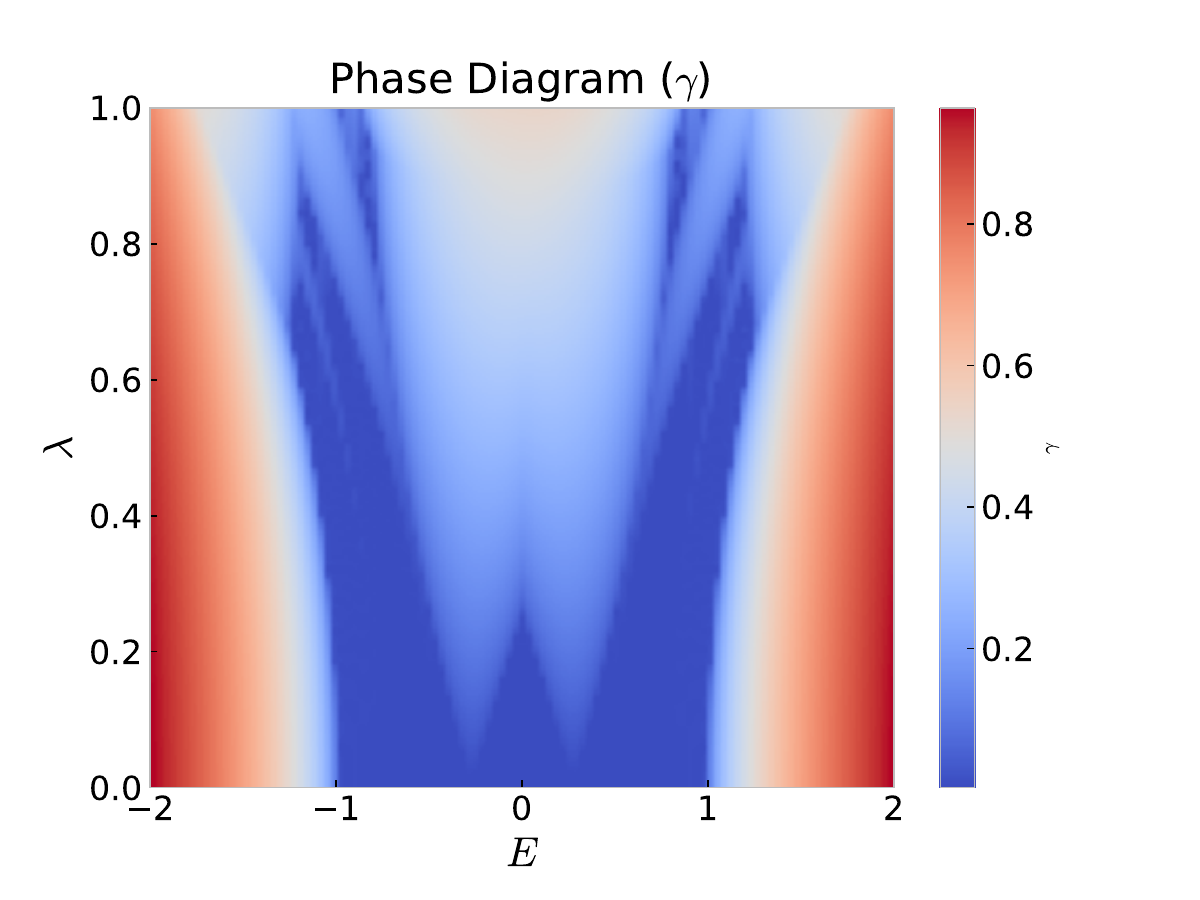}
\caption{Phase diagram of the symmetric onsite energy model. The x-axis represents the energy of the system $E$, the y-axis represents  $\lambda$, and the color bar indicates the Lyapunov exponent $\gamma$. System size $N=200$. $t=\tilde{t} = -1$.}
\label{fig:phase_diagram_symmetric}
\end{figure}

First, we examine the behavior of the Lyapunov exponent $\gamma$ for
various values of $\lambda$ as a function of the energy $E$ (see
Fig. \ref{fig:lyapunov_vs_energy_symmetric}). This plot illustrates
how the Lyapunov exponent changes with energy across different
potential strengths. The results indicate that as $\lambda$ increases, the Lyapunov exponent generally increases for most of the energy spectrum, particularly near the Fermi level. However, we observe a non-monotonic behavior at the extremes of the energy spectrum, where the Lyapunov exponent tends to decrease at both ends, namely in the ranges \( E \in (-\infty, -1.2) \) and \( E \in (1.2, +\infty) \). This suggests that the increase in localization with $\lambda$ is more pronounced around the Fermi level, while the behavior at extreme energies requires further investigation. To fully
understand it, we also present a phase diagram of the symmetric onsite
energy model in Fig. \ref{fig:phase_diagram_symmetric}. This figure
illustrates the behavior of the Lyapunov exponent $\gamma$ across a
range of energy values $E$ and potential strengths $\lambda$. As we can
see there are mobility edges, that differentiate between energies with
vanishingly small $\gamma$ and those with finite values of
$\gamma$. At very low values of $\lambda$, the states at the middle of
the spectrum have vanishingly small $\gamma$, but as $\lambda$
increases, it yields to phase diagram with two branches, i.e. at the
middle of the spectrum, and at the two edges of the spectrum Lyapunov
exponent is finite, while there are two symmetric narrow of energy
that has vanishingly small values of Lyapunov exponent.

It is important to note that our model differs from the one studied in Ref. \cite{PhysRevB.99.054211}. While Ref. \cite{PhysRevB.99.054211} focuses on the effects of off-diagonal coupling between channels, our model does not include this type of coupling. Moreover, unlike in Ref. \cite{PhysRevB.99.054211}, where an intermediate energy region exists with a combination of localized and delocalized states, our model exhibits distinct phase transitions without such mixed regions. This distinction leads to different localization properties and phase diagrams.

\begin{figure*}
\centering
\begin{subfigure}{}%
\includegraphics[width=0.33\textwidth]{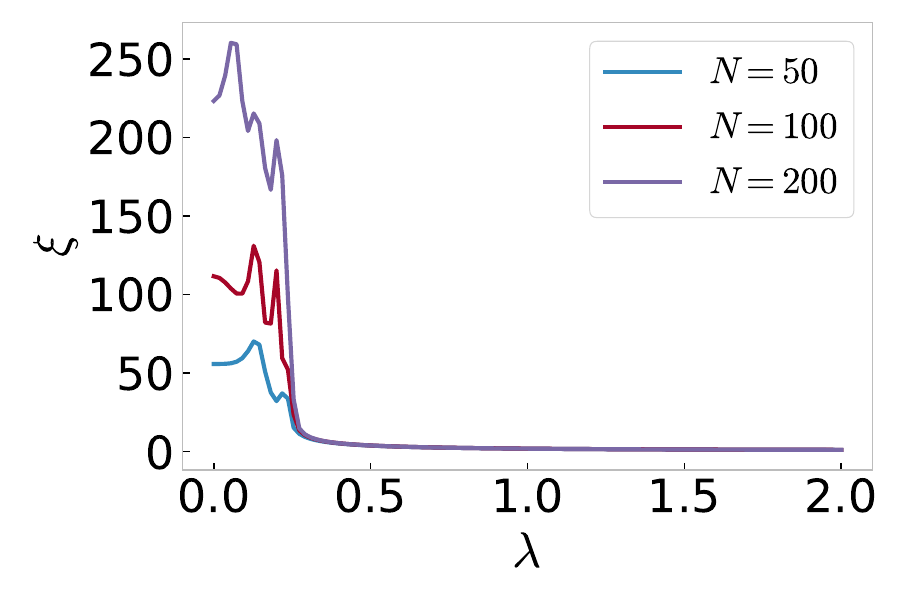}
\end{subfigure}
\begin{subfigure}{}%
\includegraphics[width=0.33\textwidth]{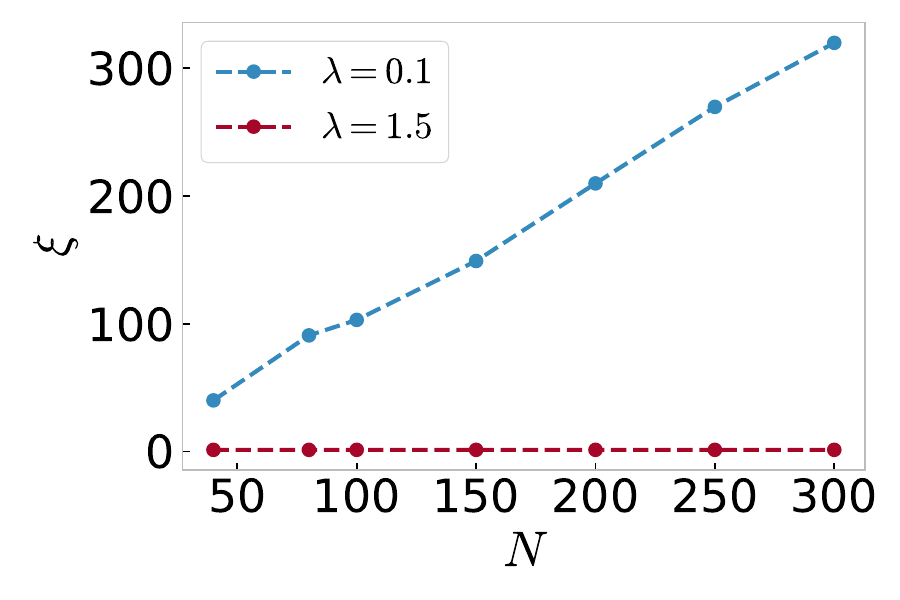}
\end{subfigure}%
\begin{subfigure}{}%
\raisebox{0.3cm}{
\includegraphics[width=0.32\textwidth]{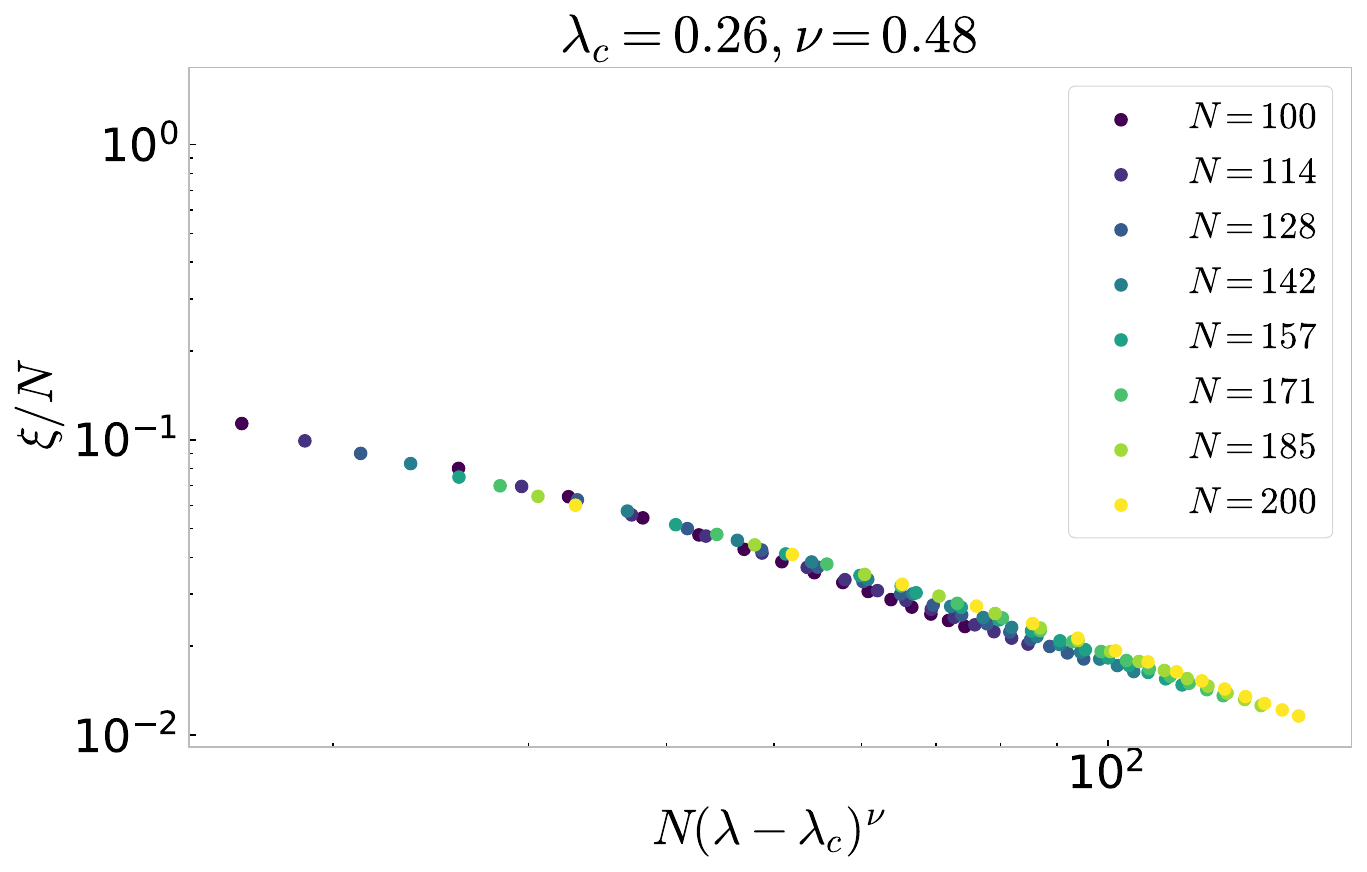}
}
\end{subfigure}%
\caption{Localization length in the symmetric onsite energy model. Left panel: Localization length $\xi$ as a function of $\lambda$ at $E =
  0$. The localization length decreases with
  increasing $\lambda$, indicating stronger localization. As we can
  see there are some oscillations of localization length in the
  delocalized phase. Middle panel: Localization length $\xi$ as a function of system size $N$ at
  $E = 0$ for specific values of $\lambda$. The
  localization length increases with system size $N$, indicating
  delocalized phase, ($\xi \sim N$) or remains constant indicating
  localized phase ($\xi \sim 1$). Right panel: Data collapse of the rescaled localization lengths $\xi/N$
  against the scaling variable $N (\lambda - \lambda_c)^{\nu}$ for $E=0$. We set $t=\tilde{t} = -1$ for all plots.}
\label{fig:xisymmetric}
\end{figure*}

We also plotted the localization length for selected system sizes $N$ as a function of $\lambda$ for $E = 0$ (see left panel of Fig. \ref{fig:xisymmetric}). At both energies, we observe a critical value of $\lambda$ beyond which the localization length vanishes, indicating a phase transition between delocalized and localized states. The precise values of these critical points and their corresponding critical parameters will be determined subsequently. To further distinguish the behavior in the delocalized and localized phases, we plot the localization length versus system size for selected values of $\lambda$ in middle panel of Fig. \ref{fig:xisymmetric}. This plot confirms that the localization length scales linearly with system size in the delocalized phase ($\xi \sim N$) and remains constant in the localized phase ($\xi \sim 1$).

To further understand the localization properties, we apply the method
of the cost function described previously in subsection
\ref{costfunction} to determine the critical exponent $\nu$ and the
critical localization length $\lambda_c$. Using the cost function approach, we perform a scaling analysis to find the optimal parameters $\lambda_c$ and $\nu$ that
minimize the cost function $\mathcal{C}_Q$. The optimal parameters $\lambda_c$ and $\nu$ are determined from the run with the lowest cost function value. The successful data collapse, as visualized by plotting the rescaled localization lengths against
the scaling variable $N (\lambda - \lambda_c)^{\nu}$, confirms the
accuracy of the obtained critical parameters (see right panel of
Fig. \ref{fig:xisymmetric}).

\subsection{Asymmetric Onsite Energy Model}
In the asymmetric onsite energy model, where $\lambda_1 \neq \lambda_2$, we introduce differing onsite potential strengths for the two channels. This asymmetry leads to a richer phase diagram and more intricate localization behavior compared to the symmetric case. By analyzing the localization length $\xi$ and phase transitions, we aim to explore how the imbalance between the two channels influences the critical properties of the system. In the following subsections, we extend this study to investigate the effect of varying inter-chain coupling $\tilde{t}$ and the system's energy levels.

\begin{figure}
\centering
\def\big{\includegraphics[width=\linewidth]{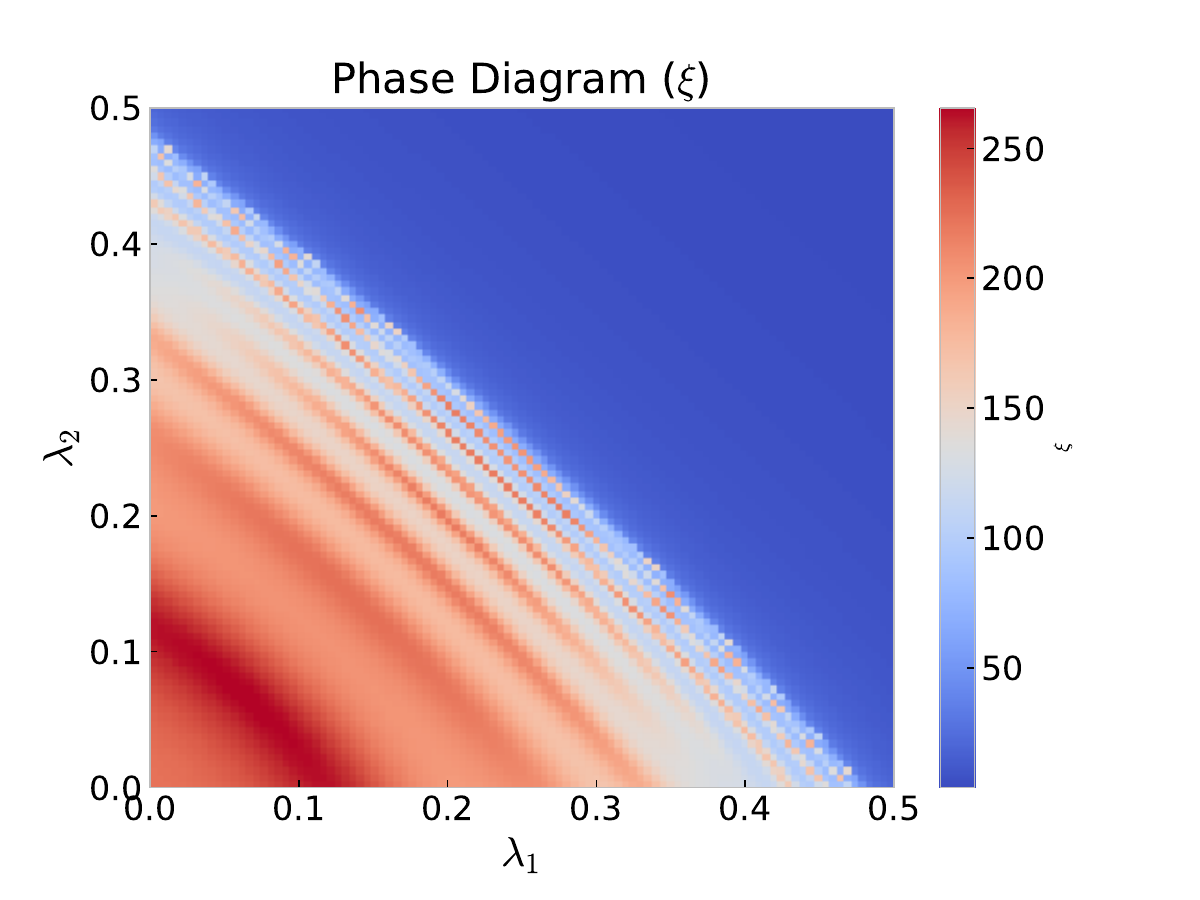}}
\def\little{\includegraphics[trim={10pt 15pt 10pt 10pt},clip,width=0.17\textwidth]{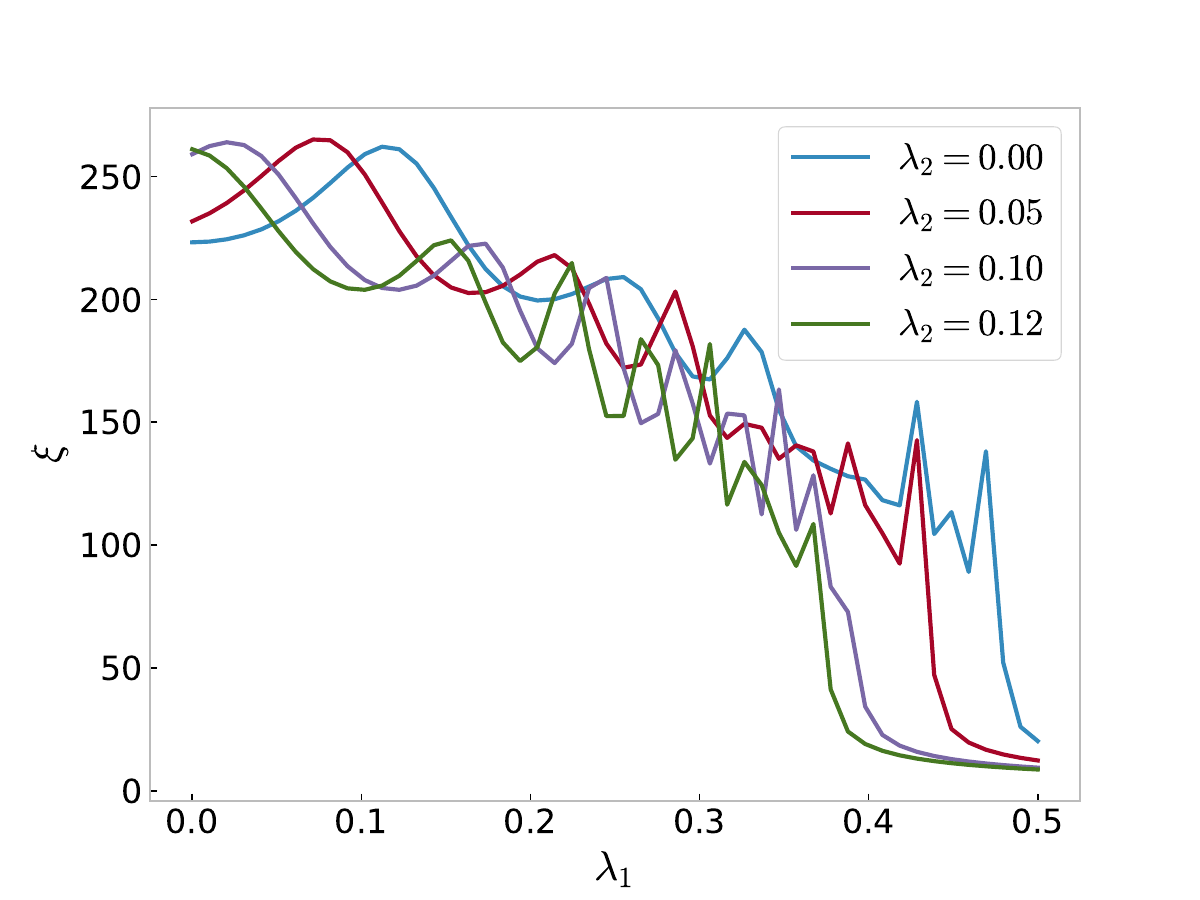}}
\stackinset{r}{63pt}{b}{100pt}{\little}{\big}
\caption{Main plot: phase diagram of the asymmetric onsite energy
  model. The x-axis represents $\lambda_1$, the y-axis represents
  $\lambda_2$, and the color bar indicates the localization length
  $\xi$. The line $\lambda_1^c + \lambda_2^c \approx 0.5$ marks the phase transition between delocalized and localized phases. Below this line, the localization length has a large value on the order of the system size, indicating delocalized states. Above this line, the localization length is effectively zero, signifying localized states. Inset plot: Localization length $\xi$ as a function of $\lambda_1$ for various values of $\lambda_2$ in the Asymmetric onsite energy model. The
  localization length decreases with increasing $\lambda_1$,
  indicating stronger localization. We set $E=0$, and system size
  $N=200$, $t=\tilde{t} = -1$.}
\label{fig:phase_diagram_asymmetric}
\end{figure}

First, we study the case of $\tilde{t}=-1$, i.e. when all hopping
amplitudes are the same and at the middle of the spectrum $E=0$. We
present a phase diagram (see Fig. \ref{fig:phase_diagram_asymmetric}),
using a scatter plot where the x-axis represents $\lambda_1$, the
y-axis represents $\lambda_2$, and the color at each data point
indicates the magnitude of the localization length $\xi$. This plot
reveals the regions of localized and delocalized phases. The phase
transition is observed around the line $\lambda_2^c + \lambda_1^c \approx\frac{1}{2}$, below which the localization length is larger,
indicating a delocalized phase (although there are some oscillatory
behavior of the localization length, see the inset plot of the
Fig. \ref{fig:phase_diagram_asymmetric}).

\begin{figure}
\centering
\begin{subfigure}{}%
\includegraphics[width=0.23\textwidth]{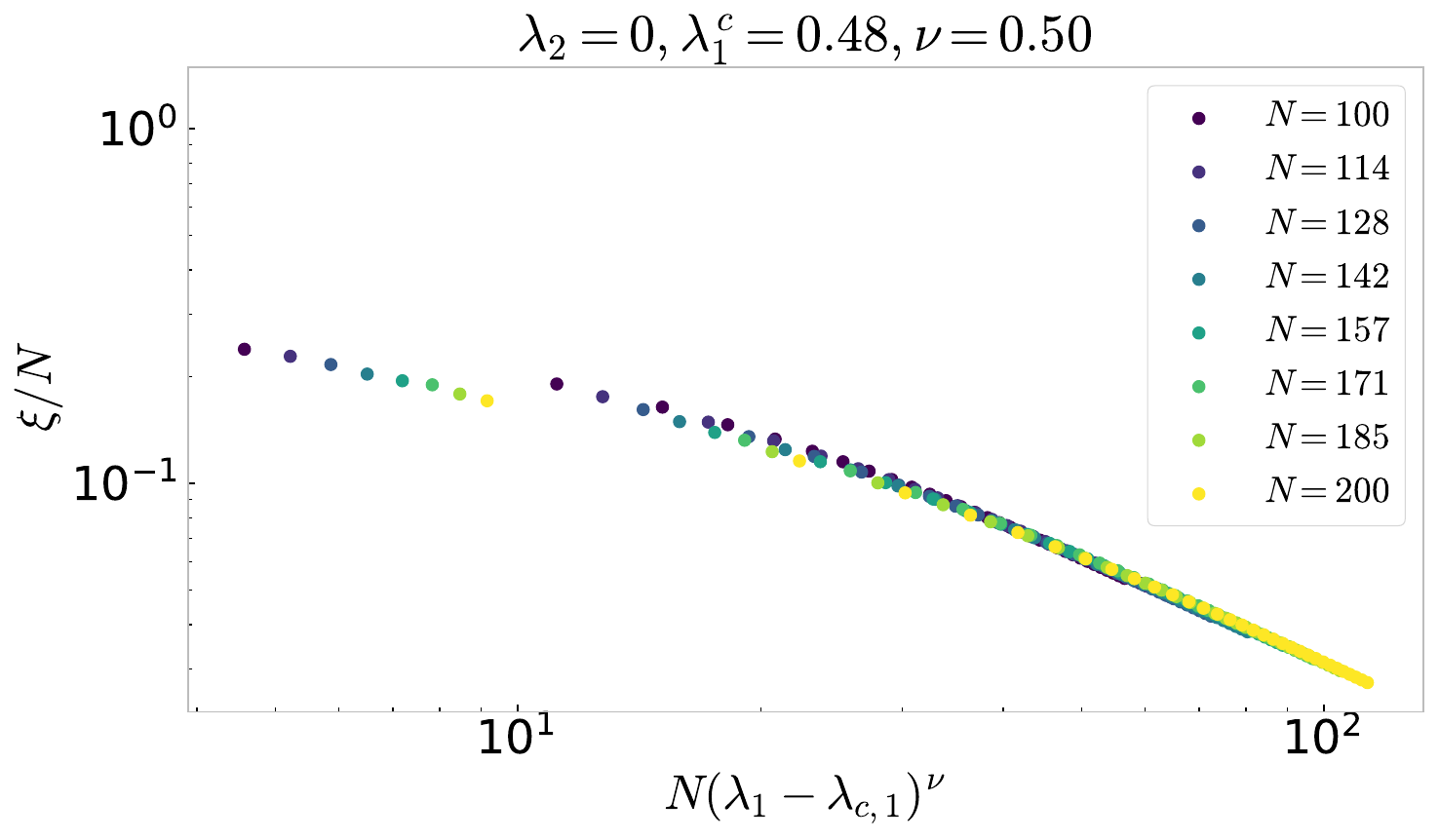}
\end{subfigure}
\begin{subfigure}{}%
\includegraphics[width=0.23\textwidth]{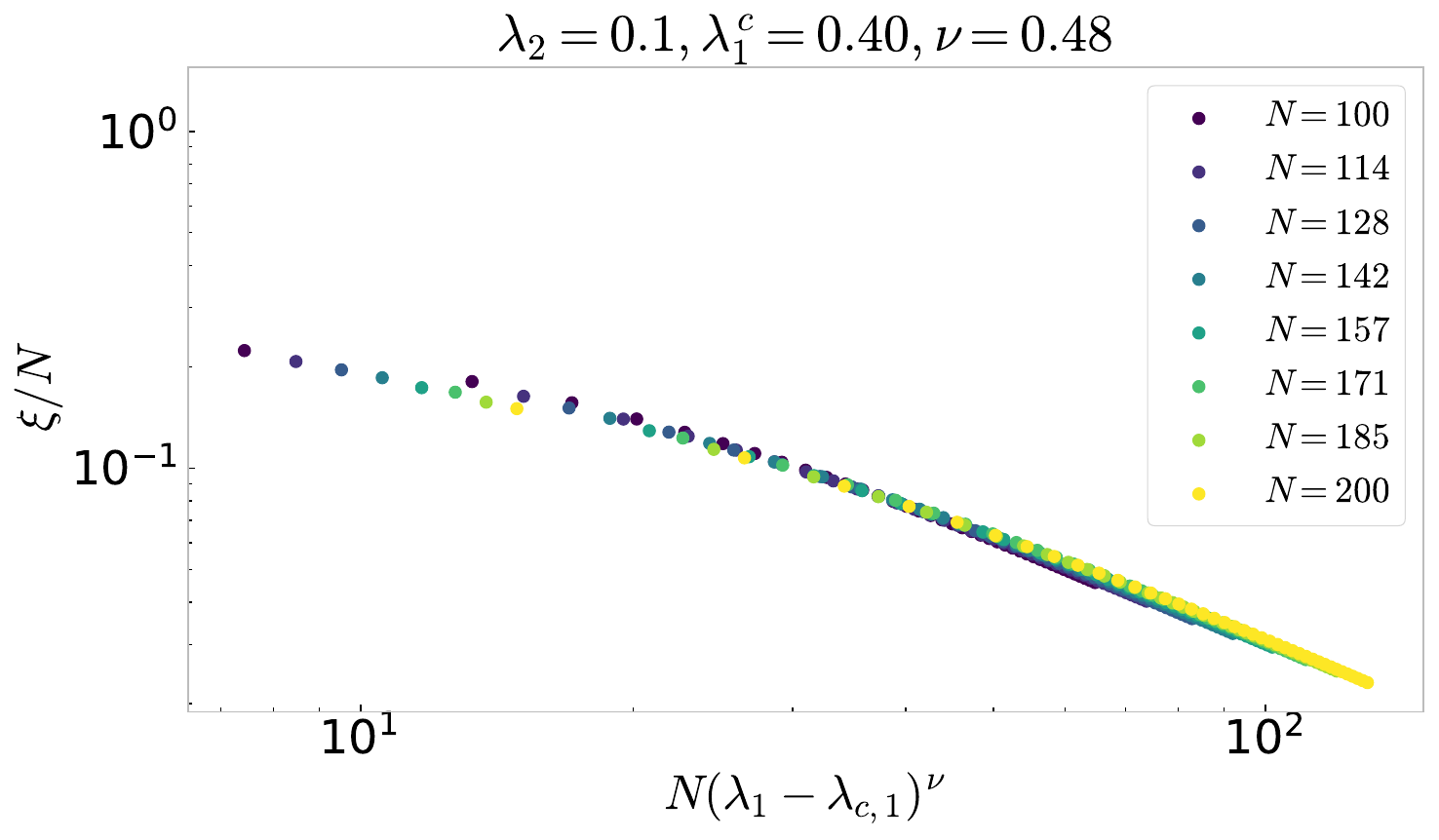}
\end{subfigure}%
\begin{subfigure}{}%
\includegraphics[width=0.23\textwidth]{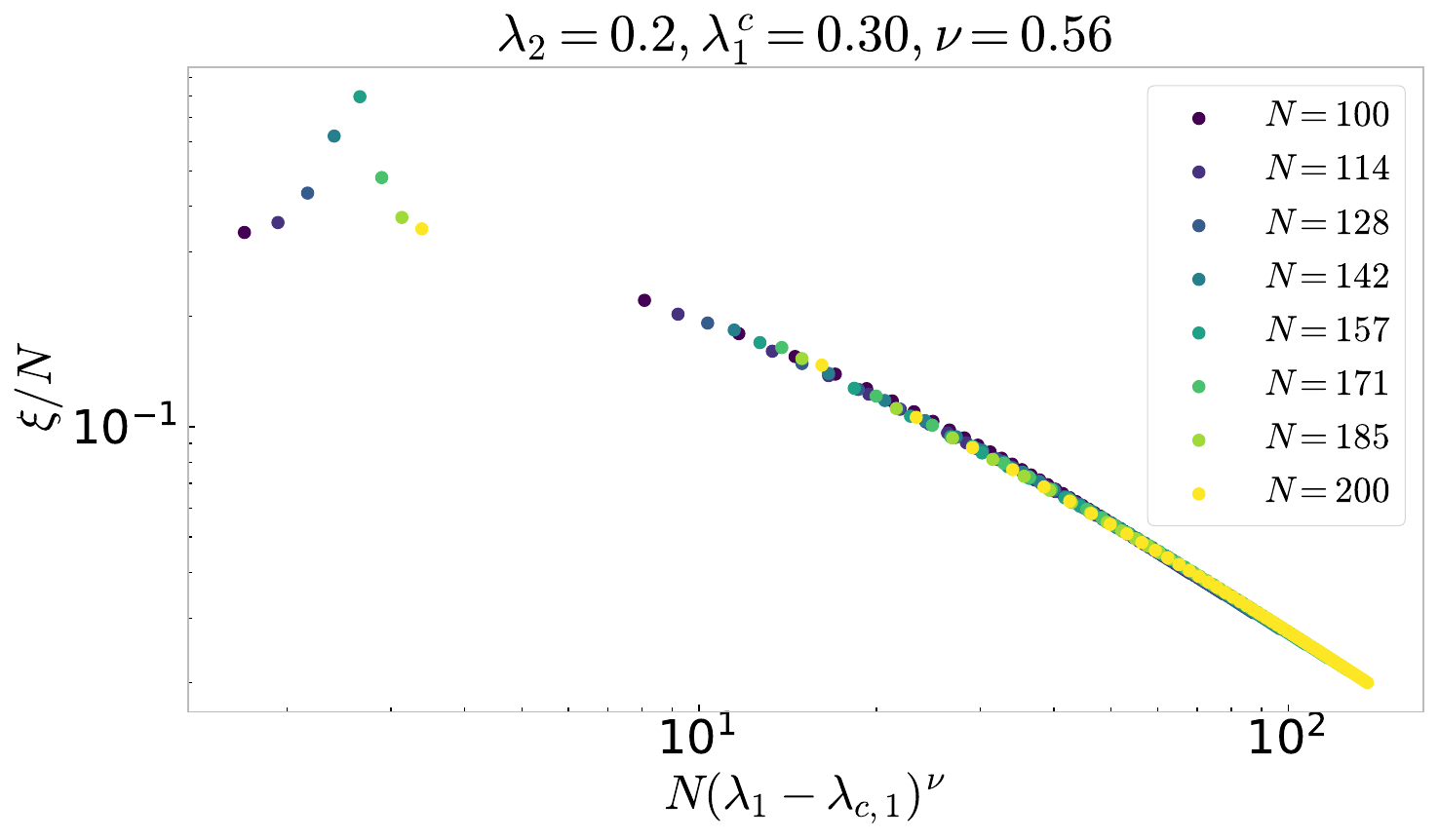}
\end{subfigure}%
\begin{subfigure}{}%
\includegraphics[width=0.23\textwidth]{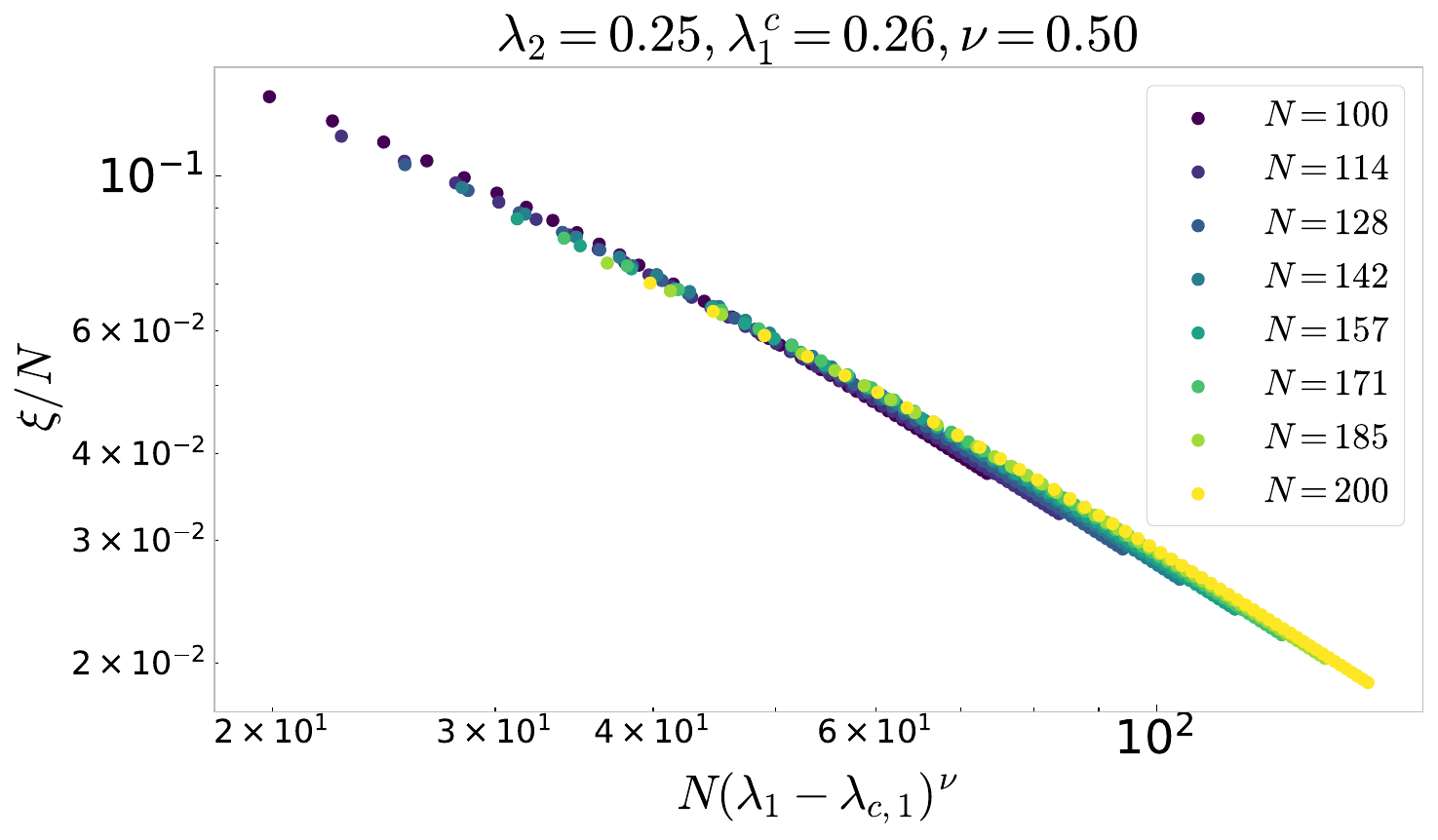}
\end{subfigure}%
\caption{Data collapse of the rescaled localization lengths $\xi/N$
  against the scaling variable $N (\lambda_1 - \lambda_c)^{\nu}$ in
  the Asymmetric onsite energy model. For some selected values of $\lambda_2$, the $\lambda_1^c$ and its corresponding $\nu$ is obtained using the cost function. We set $t=\tilde{t} = -1$ for all plots.}
\label{fig:data_collapse_asymmetric}
\end{figure}

To further characterize and locate the phase transition point at the middle of the spectrum ($E=0$), we use the cost function approach to calculate the critical potential strength $\lambda_c$ and the critical exponent $\nu$ for selected values of $\lambda_2$ (we could alternatively choose $\lambda_1$), see Fig. \ref{fig:data_collapse_asymmetric}. As shown, the values of the critical $\lambda_1$ are consistent with the phase diagram's observation of the critical line $\lambda_2 + \lambda_1 = 0.5$. The critical exponent $\nu$ is approximately $0.5$ same for all selected values of $\lambda_1$.

\subsection{Influence of Inter-Chain Coupling: $\tilde{t}$}
\begin{figure}
\centering
\includegraphics[width=\linewidth]{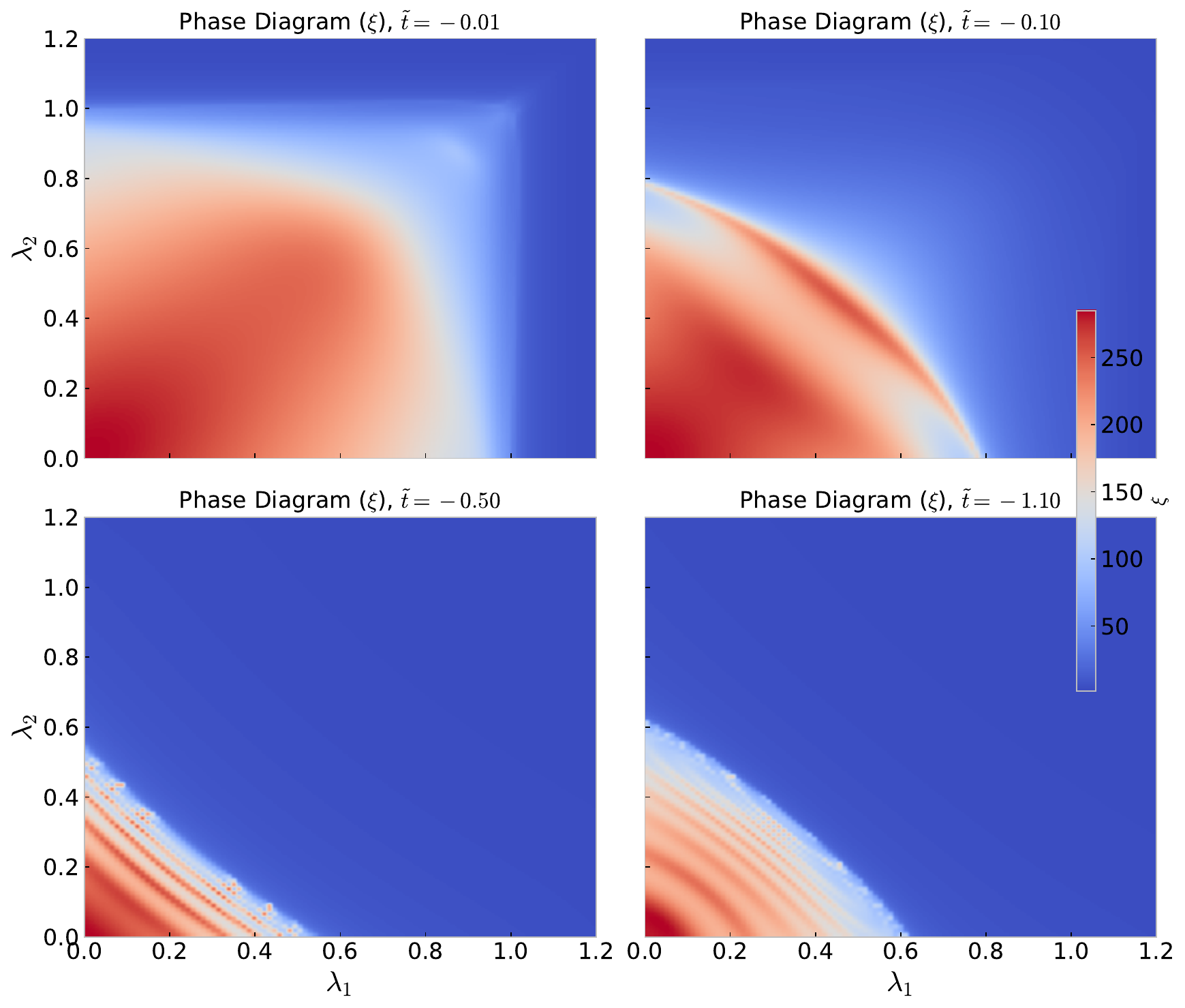}
\caption{Phase diagram of the asymmetric onsite energy model. The x-axis represents $\lambda_1$, while the y-axis represents $\lambda_2$. The color bar indicates the localization length $\xi$. Each plot corresponds to a specific value of $\tilde{t}$. The behavior of the localization length and the phase transition point is influenced by the value of $\tilde{t}$. For very small values of $\tilde{t}$, the system approaches the behavior of uncoupled chains. As $\tilde{t}$ increases, the phase space of the delocalized phase decreases. We set $N=200$, $t=-1$.}
\label{fig:tildet}
\end{figure}

In this section, we explore the effect of the inter-chain coupling strength $\tilde{t}$ on the localization length $\xi$ and the phase transition behavior of the system. The coupling term $\tilde{t}$ introduces interactions between the two channels, modifying the effective onsite energies for both chains. By varying $\tilde{t}$, we observe significant changes in the localization length $\xi$, as illustrated in Fig. \ref{fig:tildet}. The strength of the inter-chain coupling strongly influences the position of the phase transition. For very small values of $\tilde{t}$, the phase transition remains close to the decoupled case, where $\lambda_1^c = 1$ and $\lambda_2^c = 1$, similar to the $\tilde{t} = 0$ scenario. This observation makes intuitive sense, as the weak coupling between the chains does not significantly alter the effective onsite energies. However, as $\tilde{t}$ increases, the "phase space" of the delocalized region, where the system exhibits large values of $\xi$, gradually shrinks. The phase transition shifts closer to the condition $\lambda_1^c + \lambda_2^c \approx 0.5$. This suggests that stronger coupling forces the two channels to interact in such a way that the phase boundary is no longer governed by the individual critical potential strengths $\lambda_1$ and $\lambda_2$, but rather by their combined effect.

The shift of the phase transition towards $\lambda_1^c + \lambda_2^c \approx 0.5$ as $\tilde{t}$ increases can be understood by considering the inter-chain coupling's effect on the system. When $\tilde{t} \neq 0$, the effective onsite energies for both chains become nonlinear functions of $\lambda_1$ and $\lambda_2$. This nonlinearity causes the potential landscape to change, with the coupling $\tilde{t}$ effectively "blending" the potential strengths of the two channels. Instead of each chain undergoing an independent phase transition at $\lambda_1^c = 1$ and $\lambda_2^c = 1$, the system's critical behavior becomes dominated by an effective sum of the two potential strengths. The result is a critical line $\lambda_1^c + \lambda_2^c \approx 0.5$, where the combined disorder is just strong enough to induce localization across the coupled system.

\subsection{Influence of Energy levels}
\begin{figure}
\centering
\includegraphics[width=\linewidth]{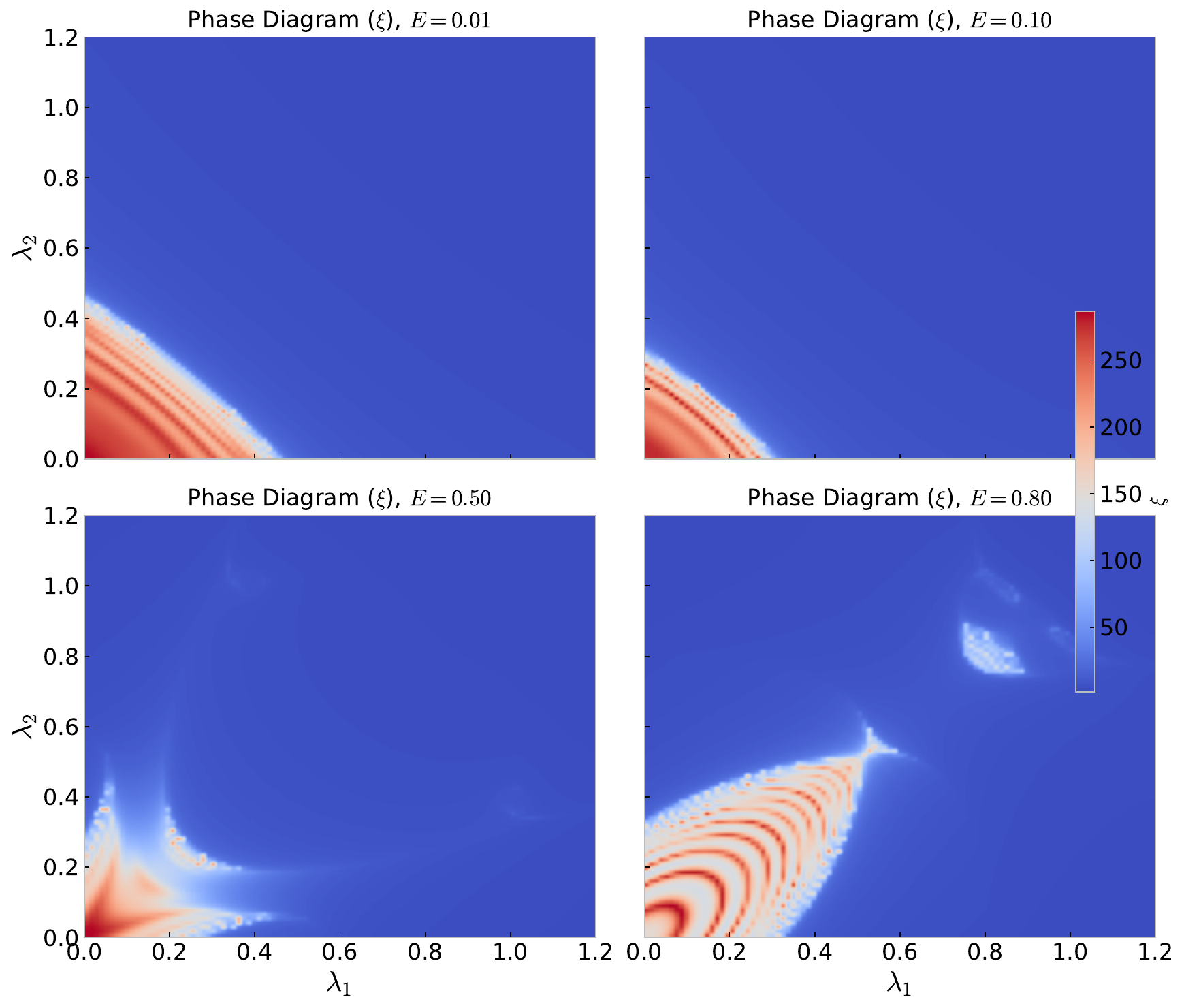}
\caption{Phase diagram for the asymmetric onsite energy model. The
  horizontal axis denotes $\lambda_1$, while the vertical axis
  indicates $\lambda_2$. The color scale represents the localization
  length $\xi$. Each subplot corresponds to a distinct value of
  $E$. The localization length and the phase transition point exhibit
  a dependence on the energy level $E$. For very small values of $E$,
  the system's behavior resembles that of a system centered around the
  mid-energy region. As the energy $E$ increases, the extent of the
  delocalized phase in the phase space diminishes. We set $N=200$, and $t=\tilde{t} = -1$.}
\label{fig:E}
\end{figure}

In this section, we investigate the effects of the energy levels of
the system. Thus far, our analysis has focused on the central region
of the energy spectrum. Now, we observe that the phase diagram is
dependent on the energy levels of the system. Specifically, at
energies near $0$, the phase diagram exhibits characteristics similar
to that at $E=0$. However, as the energy level is altered, the phase
diagram and the location of the phase boundary undergo significant
changes (See Fig. \ref{fig:E}).

\subsection{Effect of the Incommensurate Parameter $b$}

In this subsection, we study the influence of the incommensurate parameter $b$ on the localization properties of the system. While the general behavior of phase transitions remains the same, there are subtle differences in the phase diagrams for different values of $b$. The parameter $b$ controls the quasiperiodicity of the potential, and changing $b$ affects the structure of the phase boundaries and the localization length $\xi$.

We plot the localization length $\xi$ for four different values of $b$: $\frac{\sqrt{2}}{2}$, $\frac{\sqrt{5} - 1}{2}$, $\pi$, and $\frac{\sqrt{3}}{2}$. These values of $b$ are chosen to represent different types of irrational numbers, and our results show that, while the overall phase transition structure is preserved, the details of the phase diagrams, such as the width of the delocalized region, vary slightly with $b$.

\begin{figure}
	\centering
	\includegraphics[width=\linewidth]{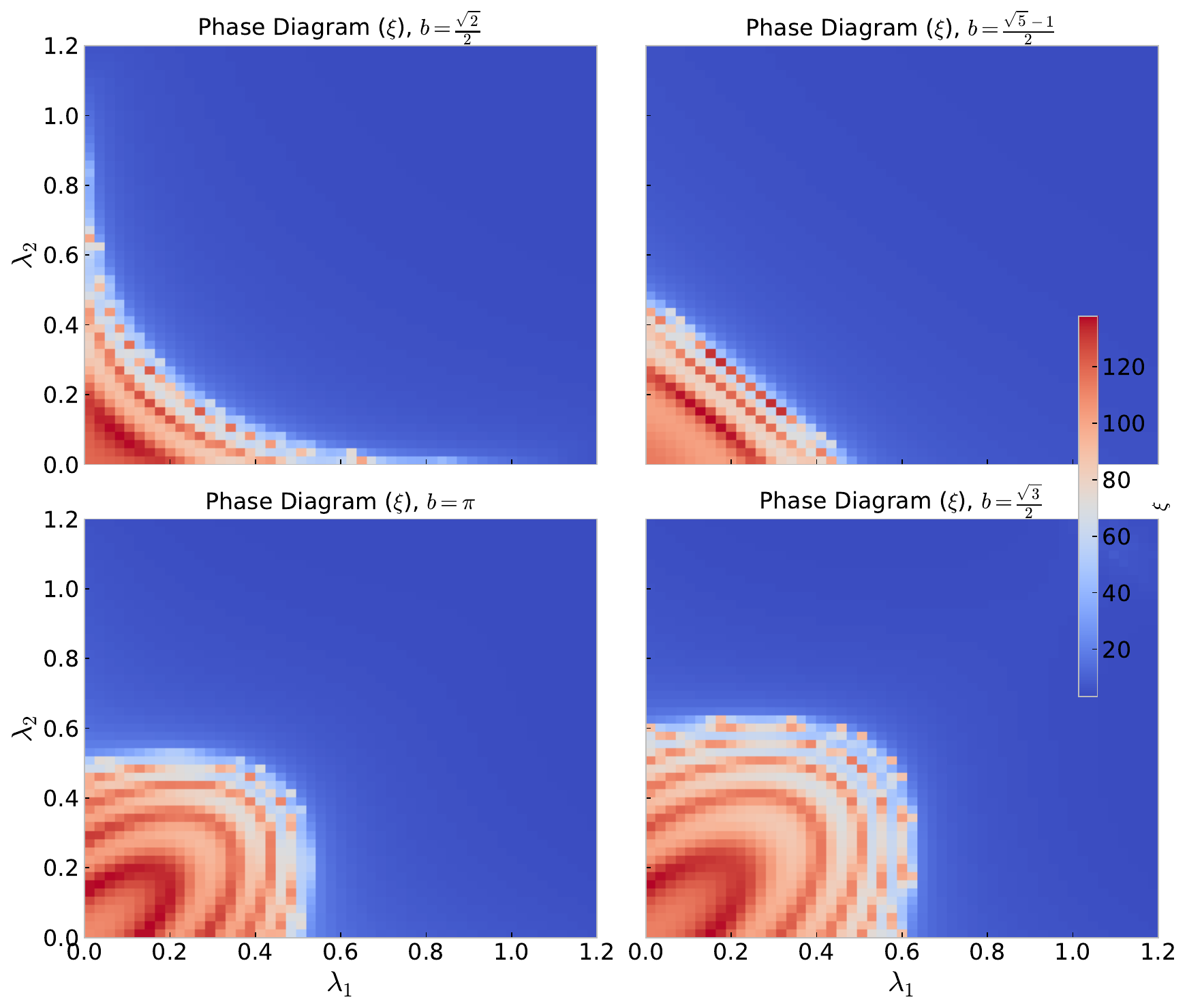}
	\caption{Phase diagram of the localization length $\xi$ for different values of the incommensurate parameter $b$: $\frac{\sqrt{2}}{2}$, $\frac{\sqrt{5} - 1}{2}$, $\pi$, and $\frac{\sqrt{3}}{2}$. The behavior of the phase transitions remains consistent across different values of $b$, though the details of the phase boundaries vary slightly with $b$. We set $N=200$, and $t=\tilde{t} = -1$.}
	\label{fig:b_comparison}
\end{figure}

As seen in Fig. \ref{fig:b_comparison}, the transition from the localized to delocalized phase is sharp for all values of $b$, but the exact location of the transition is sensitive to the choice of $b$. This sensitivity is expected, given the quasiperiodic nature of the Aubry-André model, and emphasizes the role of $b$ in shaping the localization properties of the system.

\section{Conclusion and Future Work}
\label{conclusion}

In this work, we investigated the localization properties of a quasi-one-dimensional two-channel system using the Aubry-Andr\'{e} model for both symmetric and asymmetric onsite energy configurations. By analyzing the Lyapunov exponent and localization length, we characterized the phase transitions and critical behavior of the system across various parameter regimes.

For the symmetric onsite energy model, our results demonstrated that the Lyapunov exponent increases with potential strength, leading to stronger localization. We presented phase diagrams that illustrate the behavior of the Lyapunov exponent across different energies and potential strengths. The critical exponent $\nu$ and the critical potential strength $\lambda_c$ were obtained using the cost function method for various energy levels.

In the study of the asymmetric onsite energy model, we examined the effects of differing potential strengths in the two channels. Our analysis revealed a critical line $\lambda_1^c + \lambda_2^c \approx 0.5$, that shows the phase transition between delocalized and localized phases. Additionally, we studied the effects of the inter-channel coupling parameter $\tilde{t}$ and the behavior of the localization length $\xi$. We found that for small $\tilde{t}$, the phase transition closely resembles the case where $\lambda_1 = 1$ and $\lambda_2 = 1$, while increasing $\tilde{t}$ causes the delocalized phase space to shrink, aligning more closely with the critical line $\lambda_1^c + \lambda_2^c \approx 0.5$. These findings, supported by semi-analytical explanations, provide insight into how inter-channel coupling impacts localization. We also explored the effects of varying the energy level $E$ in the system.

Furthermore, we studied different incommensurate parameters $ b $. While the general behavior of phase transitions remains the same—shifting from $ \lambda_1 = \lambda_2 = 1 $ at $ \tilde{t} = 0 $ to $ \lambda_1^c + \lambda_2^c \approx 0.5 $ for larger $ \tilde{t} $—there are subtle differences in the phase diagrams for different values of $ b $. In particular, for $ \tilde{t} > 1 $, the electron dynamics are characterized by hopping between the two channels rather than crossing between them. This results in a less pronounced phase transition, with the delocalized phase becoming almost indistinguishable. As a consequence, the system exhibits localization, which, while interesting, is less relevant for the study of localization-delocalization transitions in our model. Therefore, the focus remains on systems with $ \tilde{t} \leq 1 $, where the ability of electrons to cross between channels plays a critical role in the localization process.

In recent studies of the Aubry-André model, the critical exponent $\nu = 1$ for the localization-delocalization transition has been well established. Sinha et al. (2019) and Wei (2019) both obtained $\nu = 1$ through their respective analyses of the Kibble-Zurek mechanism and fidelity susceptibility in one-dimensional systems \cite{PhysRevB.99.094203, PhysRevA.99.042117}, confirming the scaling behavior $\xi \sim (\lambda - 1)^{-\nu}$. Further insights into boundary conditions and finite-size scaling in one-dimensional systems were provided by Thakurathi et al. (2012) and Hashimoto et al. (1992), contributing to the understanding of critical points in localization transitions \cite{PhysRevB.86.245424, YHashimoto_1992}. In contrast, our calculations yield a critical exponent of $\nu = 0.5$, which deviates from the established value of $\nu = 1$ for one-dimensional systems, suggesting potential differences in higher dimensions. While $\nu = 1$ has been consistently observed in 1D models, our result may point to distinct behavior in two-dimensional systems, where the critical exponent could differ due to the enhanced spatial degrees of freedom. This discrepancy warrants further exploration of localization transitions in 2D quasiperiodic systems, an area that remains less explored in the current literature.

The Aubry-André (AA) model, originally formulated to describe localization phenomena in quasiperiodic systems, has found a variety of experimental realizations, highlighting its relevance to modern condensed matter and cold atom research. In a study by An et al. (2021), a generalized AA model was experimentally realized using synthetic lattices of laser-coupled atomic momentum modes, demonstrating the emergence of a mobility edge, a hallmark of localization-delocalization transitions \cite{PhysRevLett.126.040603}. Similarly, Ray et al. (2018) explored drive-induced delocalization in the AA model, showing how periodic driving can lead to the formation of a mobility edge in a system initially exhibiting localized behavior \cite{ray2018drive}. Additionally, Schreiber et al. (2015) observed many-body localization in a quasirandom optical lattice, providing experimental evidence of the interplay between disorder and interactions, and paving the way for further studies on localization phenomena in interacting systems \cite{doi:10.1126/science.aaa7432}. These studies underscore the practical relevance of the AA model in experimental settings, and there is ample room to extend these findings to more complex systems. In particular, the two-channel AA model introduced here can also be generated and utilized in experiments by generalizing the approaches discussed above. This extension opens up additional applications, potentially broadening our understanding of localization phenomena and offering new insights into higher-dimensional systems.

In this context, the asymmetric two-channel AA model introduced in this work represents a significant advancement, offering both theoretical and practical relevance. Its ability to model systems with asymmetric disorder or coupling, such as multilayered materials or quantum wires with spin-orbit coupling, extends the applicability of the AA model to more complex scenarios. Furthermore, this model can be used to investigate the impact of asymmetry on quantum transport properties, providing new insights into disordered systems, quantum coherence, and transport in both mesoscopic and cold atom setups. By generalizing the approaches discussed in the aforementioned studies, the two-channel AA model can also be generated and utilized in experiments, opening up additional applications and potentially broadening our understanding of localization phenomena in higher-dimensional and more intricate systems.

For future work, it would be beneficial to investigate the full energy spectrum of the asymmetric model in greater detail and explore other models with intrinsic mobility edges to examine how different forms of disorder affect localization. Additionally, a promising avenue for future research is to introduce tunable correlated randomness in place of the Aubry-Andr\'e potential, allowing for a more general study of the interplay between correlated disorder and localization. These directions aim to deepen our understanding of complex localization phenomena in low-dimensional systems.

\section*{Data Availability Statement}
The datasets generated during and/or analyzed during the current study
are available from the corresponding author on reasonable request.

\acknowledgments
This work was supported by the University of Mazandaran.

\bibliographystyle{apsrev4-2.bst}
\bibliography{/home/cmp/Dropbox/physics/Bib/reference.bib}
\end{document}